\documentclass[preprint,eqsecnum,aps]{revtex4}
\input{epsf}
\usepackage{graphicx}%
\usepackage{dcolumn}
\usepackage{amsmath}

\makeatletter

\def\2pap{2\pi\alpha^\prime}

\def\beq{\begin{eqnarray}}
\def\eeq{\end{eqnarray}}
\def\jhep#1#2#3{JHEP {\bf #1} (#2) #3}
\def\np#1#2#3{Nucl. Phys. {\bf B#1} (#2) #3}
\def\pl#1#2#3{Phys. Lett. {\bf B#1} (#2) #3}
\def\prl#1#2#3{Phys. Rev. Lett.{\bf #1} (#2) #3}
\def\pr#1#2#3{Phys. Rev. {\bf D#1} (#2) #3}

\begin{document}

\rightline{hep-th/0212009}
%\title{Intersecting D-branes under Infinite Lorentz Boost}
\title[Short Title]{Supersymmetric Boost on Intersecting D-branes}

\author{Jin-Ho Cho}\email{jhcho@taegeug.skku.ac.kr} \author{Phillial Oh}\email{ploh@newton.skku.ac.kr}
 %\thanks{Also at Physics Department, XYZ University.}
 
 \affiliation{BK21 Physics Research Division and Institute of Basic Science\\
 Sungkyunkwan University, Suwon 440-746, Korea}

\date{\today}% It is always \today, today, but you may specify any date with \date.

\begin{abstract}
We study the effect of the Born-Infeld electric field on the supersymmetric configuration of various composite D-branes. We show that the generic values of the electric field do not affect the supersymmetry but, as it approaches $1/2\pi\alpha'$ keeping the magnetic field finite, various combinations of the magnetic fields allow up to 8 supersymmetries. We also explore the unbroken supersymmetries for two intersecting D-strings which are in uniform or relative motion. For a finite uniform Lorentz boost, 16 supersymmetries are guaranteed only when they are parallel. For an infinite one, 8 supersymmetries are preserved only when both the D-strings are oriented to the forward or backward direction of the boost. Under a finite relative boost, 8 supersymmetries are preserved only when the intersecting angle is less than $\pi/2$ and the intersecting point moves at the speed of light. As for an infinite relative boost, 8 supersymmetries are preserved regardless of the values of the intersecting angle.
\end{abstract}

% PACS, the Physics and Astronomy Classification Scheme.
%Use showkeys class option if keyword
                              %display desired
\maketitle

\section{Introduction}

Supersymmetric configurations of multiple D-branes have received intensive studies. 
A D0-D(2$k$)-brane system is supersymmetric when $k$ is even valued \cite{johnson,polchinski}. This configuration is U-dual to two D-, NS-, or M-branes intersecting at right angles. More generally, two D$k$-branes related by $SU(k)$ rotations preserve some supersymmetries unbroken \cite{berkooz}. Especially even for odd values of $k$, D0-D(2$k$)-brane system can be supersymmetric in the presence of suitable background Born-Infeld (BI) magnetic fields \cite{park,witten}. 

Along this line of research, the supertube (a tubular D2-brane) \cite{townsend} attracts our interests in many respects. Firstly, it preserves 8 supersymmetries in spite of the tubular shape (see \cite{lee} for its matrix description). It is self supported against collapse by the Poynting momentum around the tube circumference, and this is provided by the BI electric and magnetic fields over the tube worldvolume. Secondly, it opens up the study on the supersymmetric D-$\overline{\mbox{D}}$ brane systems \cite{bak}. The circular cross section is deformable to any different shape including the elliptic one \cite{mateos,cho}. There is no tension along the cross section. An extreme deformation of the elliptic supertube results in a D2-$\overline{\mbox{D}2}$ pair, and there is no tachyonic mode on the interstring running between them \cite{ohta}.  
In all these supersymmetric configurations, the `critical' electric field ($E=1/2\pi\alpha'$) plays a very important role \footnote{critical in the sense that Born-Infeld action vanishes when magnetic fields are absent.}. Especially when we T-dualize these systems along the direction of the electric field, we obtain intersecting D-strings moving at the speed of light \cite{oh,lunin}. One could imagine the interplay between the intersecting angle of the D-strings and their speed to make the configuration supersymmetric. 

The purpose of this work is to understand the role of the electric field in the supersymmetric configuration of composite D-branes. We shall work in the context of the open strings living on these composite D-branes, that is, examine the conserved spacetime supercharges on the open-string worldsheet. This method was adopted in Ref. \cite{johnson,polchinski,berkooz,park,witten} to check the supersymmetries for the multiple D-branes. It is a more direct way of checking the spacetime supersymmetries than other methods used in the effective theories like the BI world volume theory or the supergravity theory.    

This paper is organized as follows. In Sec. \ref{sec2}, we specify the way to give the BI electric and magnetic fields over the D($2k$)-brane world volume. The basic idea is to take T-dualities on a tilted D$k$-brane in motion. We obtain the relations between the BI field strengths and the geometrical data on the tilting and Lorentz boost. 

In Sec. \ref{sec3}, we check the supersymmetry for various D0-D$(2k)$ brane systems dressed in the BI electric and magnetic fields. We show that for generic (`noncritical') values of the electric field, the number of the preserved supersymmetries is exactly the same as that of the system with the magnetic fields only. 

In Sec. \ref{sec4}, we work out the cases where the electric field takes the critical value. It is shown that various systems can have 8 supersymmetries at most. 

In Sec. \ref{sec5}, we study how Lorentz boost affects the supersymmetries for two D-strings intersecting at various angles. These systems are T-dual to several D2-$\overline{\mbox{D}2}$ or D2-D2 pairs depending on the intersecting angles. We show that 8 supersymmetries are preserved when both D-strings are tilted in the forward direction or both in the backward direction with respect to the boost. Otherwise, no supersymmetry is preserved. 

In Sec. \ref{sec6}, we consider the cases where two tilted D-strings are in relative motion with their mutual distance fixed. In the comoving frame of one D-string, we allow only the transverse motion of the other D-string. When the intersecting point moves at the speed of light, 8 supersymmetries are preserved only when the intersecting angle is less than $\pi/2$. When one D-string moves at the speed of light, 8 supersymmetries are preserved regardless of the intersecting angle. 

In Sec. \ref{sec7}, we conclude the paper with some remarks on various U-dual configurations and comments on future works.

\section{Dressing D$(2k)$-brane with the BI electric and magnetic fields}\label{sec2}

\subsection{Some Kinematics}

We would like to study how Lorentz boost affects the geometric configuration of a D-string lying on a plane, say, $(x,y)$-plane. Suppose it is tilted by an angle $\phi$ with respect to the coordinate $x$. Under the following boost (with boost parameter $\gamma$) along $x$-axis, 
\begin{eqnarray}
\left(
\begin{matrix} 
t'\cr x'\cr y'
\end{matrix}\right)
=\left(\begin{matrix} \cosh{\gamma}& \sinh{\gamma} & 0\cr
\sinh{\gamma} & \cosh{\gamma} & 0\cr
0 & 0 & 1
\end{matrix}
\right)
\left(\begin{matrix} t \cr x \cr y  
\end{matrix}
\right)
\end{eqnarray}
the tilting angle becomes larger;
\begin{eqnarray}\label{angle}
\tan{\phi'}=\tan{\phi}\cosh{\gamma}.
\end{eqnarray}
The velocity of the tilted D-string is given by
\begin{eqnarray}\label{velocity}
\frac{dx'}{dt'}=\tanh{\gamma}.
\end{eqnarray}

Let us consider the boundary condition of the open strings living on a tilted D-string in motion. It is convenient to introduce coordinates $(\bar{x},\bar{y})$
adapted for the D-string at rest. It is related with the coordinates $(x,y)$ as
\begin{eqnarray}\label{rel1}
\left(\begin{matrix} x\cr y
\end{matrix}
\right)
=\left(\begin{matrix} \cos{\phi}& -\sin{\phi} \cr
\sin{\phi} & \cos{\phi}
\end{matrix}
\right)
\left(\begin{matrix} \bar{x}\cr \bar{y}
\end{matrix}
\right),
\end{eqnarray}
so that the D-string lies along the $\bar{x}$-axis. The boundary conditions of the open strings 
\begin{eqnarray}
\partial_{\sigma}\bar{x}\vert_{\sigma=0,\pi}=0, \quad 
\partial_{\tau}\bar{y}\vert_{\sigma=0,\pi}=0
\end{eqnarray}
can be recasted in the boosted frame as
\begin{eqnarray}\label{bc}
&&\partial_{\sigma}t'\vert\cosh{\gamma}-\partial_{\sigma}x'\vert\sinh{\gamma}=0,\nonumber\\
&-&\partial_{\sigma}t'\vert\sinh{\gamma}\cos{\phi}+\partial_{\sigma}x'\vert\cosh{\gamma}\cos{\phi}+\partial_{\sigma}y'\vert\sin{\phi}=0,\nonumber\\
&&\partial_{\tau}t'\vert\sinh{\gamma}\sin{\phi}-\partial_{\tau}x'\vert\cosh{\gamma}\sin{\phi}+\partial_{\tau}y'\vert\cos{\phi}=0,
\end{eqnarray}
where we omitted the subscript ``${\sigma=0,\pi}$" on each vertical line $\vert$.

\subsection{T-duality}

T-duality on the tilted D-string will generate a D2-brane with a constant magnetic field over its world volume. Let us see the effect of T-duality on a tilted D-string in motion. T-duality along $x'$-direction switches the Neumann and the Dirichlet boundary condition as follows;
\begin{eqnarray}
&&\partial_{\sigma}x'\vert=0\Longrightarrow \partial_{\tau}\tilde{x}\vert=0,\nonumber\\
&&\partial_{\tau}x'\vert=0\Longrightarrow \partial_{\sigma}\tilde{x}\vert=0,
\end{eqnarray} 
where we use tilde to denote the dual coordinates. 
Those boundary conditions in (\ref{bc}) is dualized to
\begin{eqnarray}
&&\left[\partial_{\sigma}\tilde{t}-\tanh{\gamma}\,\,\partial_{\tau}\tilde{x}\right]\vert=0,\nonumber\\
&&\left[\partial_{\sigma}\tilde{y}+(\cosh{\gamma}\,\,\tan{\phi})^{-1}\,\,\partial_{\tau}\tilde{x}\right]\vert=0,\nonumber\\
&&\left[\partial_{\sigma}\tilde{x}-\tanh{\gamma}\,\,\partial_{\tau}\tilde{t}-(\cosh{\gamma}\,\,\tan{\phi})^{-1}\,\,\partial_{\tau}\tilde{y}\right]\vert=0,
\end{eqnarray}
which are nothing but the boundary conditions of open strings living on a D2-brane with the following BI field strength over its world volume;
\begin{eqnarray}\label{bifield}
2\pi\alpha'F=\tanh{\gamma}\,\,d\tilde{t}\wedge d\tilde{x}-(\cosh{\gamma}\,\,\tan{\phi})^{-1}\,\,d\tilde{x}\wedge d\tilde{y}.
\end{eqnarray}
The electric and the magnetic fields on the D2-brane can be understood as the fundamental strings arrayed along $\tilde{x}$-direction and D0-branes uniformly melted over the D2-brane, as in the case of the supertube \cite{townsend}.

We observe here that the magnetic field is inversely proportional to the slope of the D-string measured in $(x',y')$-frame. For an infinite Lorentz boost, i.e., when $\gamma\rightarrow \infty$, the D-string becomes almost parallel to $y'$-axis and the magnetic field vanishes unless $\phi\rightarrow0$. At the same time, the electric field becomes `critical' in the sense that the BI action vanishes in the limit. On the other hand, the magnetic field diverges when $\phi=0$ and the boost parameter $\gamma$ is finite, that is, $(x,y)$-frame and $(\bar{x},\bar{y})$-frame coincide. Each end point of the string is frozen to a point that can be regarded as a D$0$-brane. The momentum flow is dualized to the net number of those open strings ending on the D$0$-brane. In order to obtain a finite BI field distribution over a D$2$-brane in the limit $\gamma\rightarrow\infty$, we let simultaneously $\phi\rightarrow0$ keeping $2\pi\alpha' F_{12}=-(\cosh{\gamma}\,\,\tan{\phi})^{-1}$ finite.

\subsection{D$(2k)$-brane with the Electric- and the Magnetic Fields over its World-Volume}

We generalize the procedure discussed in the previous subsection to get a D$(2k)$-brane with the electric and the magnetic fields over its world-volume. We first consider a D$k$-brane which is tilted in each $(x_{2i-1}, x_{2i})$-plane by an angle $\phi_i$ with respect to $x_{2i-1}$-axis for $i=1,\cdots,k$. The relation between these coordinates $(x_{2i-1}, x_{2i})$ and the coordinates $(\bar{x}_{2i-1}, \bar{x}_{2i})$ adapted to D$k$-brane is the same as the relation (\ref{rel1}) but with appropriate subscripts $i$ understood on the coordinates and the angle. In order to produce the electric field over the final configuration, we boost the system and take a T-duality. Without loss of generality, the boost direction can be taken along $x_1$-axis. Taking T-dualities along $x'_1$-,$x_3$-,$\cdots,x_{2k-1}$-axes result in the following boundary conditions for the open strings living on a D$(2k)$-brane;
\begin{eqnarray}
&&\left[\partial_{\sigma}\tilde{t}-\tanh{\gamma}\,\,\partial_{\tau}\tilde{x}_1\right]\vert=0,\nonumber\\
&&\left[\partial_{\sigma}\tilde{x}_2+(\cosh{\gamma}\,\,\tan{\phi_1})^{-1}\,\,\partial_{\tau}\tilde{x}_1\right]\vert=0,\nonumber\\
&&\left[\partial_{\sigma}\tilde{x}_1-\tanh{\gamma}\,\,\partial_{\tau}\tilde{t}-(\cosh{\gamma}\,\,\tan{\phi_1})^{-1}\,\,\partial_{\tau}\tilde{x}_2\right]\vert=0,\nonumber\\
&&\left[\partial_{\sigma}\tilde{x}_{2i}+\cot{\phi_i}\,\,\partial_{\tau}\tilde{x}_{2i-1}\right]\vert=0,\nonumber\\
&&\left[\partial_{\sigma}\tilde{x}_{2i-1}-\cot{\phi_i}\,\,\partial_{\tau}\tilde{x}_{2i}\right]\vert=0,\quad (i=2,\cdots,k).
\end{eqnarray}
These boundary conditions imply the BI field strength over the D$(2k)$-brane;
\begin{eqnarray}\label{NS}
2\pi\alpha'F=\tanh{\gamma}\,\,d\tilde{t}\wedge d\tilde{x}_1-(\cosh{\gamma}\,\,\tan{\phi_1})^{-1}\,\,d\tilde{x}_1\wedge d\tilde{x}_2-\sum\limits_{i=2}^{k}\cot{\phi_i}\,\,d\tilde{x}_{2i-1}\wedge d\tilde{x}_{2i}.
\end{eqnarray}   
From here on we denote $2\pi\alpha'F_{01}\equiv E$, and $2\pi\alpha'F_{2i-1,2i}\equiv B_i$, that is, we work in the string unit $2\pi\alpha'\equiv1$. Note also that our parameters $\phi_i$ are different from the parameters $2\pi v_i$ used in \cite{witten} by $\pi/2$, that is, $2\pi v_i=\phi_i+\pi/2$.

\section{Unbroken Supersymmetries}\label{sec3}

In this section, we examine the unbroken supersymmetries for the D0-D$(2k)$ brane system in the presence of the BI field strength as in Eq. (\ref{NS}). The supersymmetry preserved by a single Dp-brane is given by the sum of the left- and right-movers on the string world sheet, $Q_{\alpha}+(\beta_p^\bot \tilde{Q})_\alpha$  \cite{polchinski}. Here 
\begin{eqnarray}
\beta_p^\bot=\prod\limits_{m=p+1}^9 \beta^m,
\end{eqnarray}
and $\beta^m=\Gamma^m\Gamma$ is the spacetime parity operator on the world sheet of the open string. In particular in the presence of the field strength like (\ref{NS}), the conserved charge becomes
\begin{eqnarray}\label{Q2k}
&&Q_{\alpha}+\left(\prod\limits_{i=1}^k \beta^{2i-1}\bar{\beta}^{2i}\right)\,\left(\beta_p^\bot \tilde{Q}\right)_\alpha,\nonumber\\
&&\quad \bar{\beta}^2=\rho(\gamma)\rho(\phi_1)\beta^2\rho(-\phi_1)\rho(-\gamma),\nonumber\\
&&\quad \bar{\beta}^{2i}=\rho(\phi_i)\beta^{2i}\rho(-\phi_i),\quad (i=2,\cdots k)
\end{eqnarray}
where $\rho(\gamma)=\exp{\left(i\gamma\Sigma^{01}\right)}$ and $\rho(\phi_i)=\exp{\left(i\phi_i\Sigma^{2i-1,2i}\right)}$ with the Lorentz rotation elements $\Sigma^{\mu\nu}=-\frac{i}{4}[\Gamma^\mu,\,\Gamma^\nu]$ of $SO(2k, 1)$ in the spinor representation. As for a D0-D$(2k)$ brane system, the unbroken supersymmetries are the intersection of the following two sets; $Q_{\alpha}+(\beta_0^\bot \tilde{Q})_\alpha$ and (\ref{Q2k}). They are in one-to-one correspondence with the spinors left invariant under
\begin{eqnarray}\label{operator}
(\beta_0^\bot)^{-1}\left(\prod\limits_{i=1}^k \beta^{2i-1}\bar{\beta}^{2i}\right)\left(\beta_p^\bot\right)&=&
\rho(\gamma)\left[\prod\limits_{j=1}^k\rho(-2\phi_j)\right]\rho(-\gamma)\\
&=&\exp{\left(-2i\sum\limits_{j=2}^k\phi_jS_j\right)}\left[\cos{\phi_1}-2i\exp{\left(\gamma\Gamma^0\Gamma^1\right)}S_1 \sin{\phi_1}\right],\nonumber
\end{eqnarray}  
where the operators, $2S_j\equiv -i\Gamma^{2j-1}\Gamma^{2j}$, take eigenvalues $2s_j=\pm1$.

Let $\epsilon$ be the eigenspinor in ${\bf 16}$ with the eigenvalue $1$. 
When $\gamma<\infty$, namely, when $\vert E\vert < 1$ in the string unit, the eigenspinor equation can be recasted into 
\begin{eqnarray}\label{eigenspinor}
\left[\prod\limits_{j=1}^k\rho(-2\phi_j)\right]\,\,\tilde{\epsilon}=\exp{\left(-2i\sum\limits_{j=1}^k\phi_jS_j\right)}\,\,\tilde{\epsilon}=\tilde{\epsilon},
\end{eqnarray}
with $\tilde{\epsilon}\equiv\rho(-\gamma)\,\,\epsilon$. Therefore the equation becomes the same as that of the system with the magnetic field only and the supersymmetry analysis will be the same as that performed in Ref. \cite{witten}. 

Let us summarize the results briefly. The D0-D2 system preserves 16 supersymmetries when $\phi_1\rightarrow 0$, that is when $B_1=-\cot{\phi_1}\sqrt{1-E^2}\rightarrow\mp\infty$. In this limit, the system effectively becomes D0-D0 system \cite{seiberg} but possibly with the electric field ($\vert E\vert < 1$) over two dimensional array of D0 particles. 

The D0-D4 system preserves supersymmetries when $\phi_1=\pm\phi_2$ (8 supersymmetries for the generic values of $B_1=\pm B_2\sqrt{1-E^2}$ and 16 supersymmetries for the divergent values under which the system becomes effectively D0-D0 system.). Especially when $E\rightarrow 0$, the 8-supersymmetric case corresponds to the (anti-)self-dual ($B_1=\pm B_2$) field configuration in Euclidean 4 dimensions. In the presence of the electric field, the fields are no longer (anti-)self-dual and are defined in Lorentzian (4+1)-dimensions but still 8 supersymmetries are preserved. 

As for the D0-D6 system, generic values of $\phi_i$'s ($i=1, 2, 3$) satisfying $2\sum\limits_{j=1}^3\phi_js_j=0$, (say $\phi_1+\phi_2+\phi_3=0$), preserve 4 supersymmetries. This is when $\vert B_i\vert <\infty$ and $\vert E\vert<1$ but $B_1(B_2+ B_3)+\sqrt{1-E^2}B_2B_3=\sqrt{1-E^2}$ is satisfied. Therefore, with all fields vanishing, the system does not preserve any supersymmetry. When one of $\phi_i$'s is zero, corresponding magnetic field diverges and the system effectively reduces to D0-D4 system preserving 8 supersymmetries. When all $\phi_i$'s vanish, it becomes effectively D0-D0 system preserving 16 supersymmetries. In all cases, noncritical values of the electric field $\vert E\vert<1$ do not disturb the supersymmetries.  

The D0-D8 system preserves 2 supersymmetries for the generic values of $\phi_i$'s ($i=1, 2, 3, 4$) satisfying $2\sum\limits_{j=1}^4\phi_js_j=0$. 
When one of $\phi_i$'s is zero, corresponding magnetic field diverges and the system effectively reduces to a D0-D6 system preserving 4 supersymmetries. When two of $\phi_i$'s are zero, the system becomes effectively D0-D4 system preserving 8 supersymmetries. When all $\phi_i$'s vanish, the system corresponds to D0-D0 system preserving 16 supersymmetries. Like the other cases, noncritical values of the electric field $\vert E\vert<1$ do not change the preserved supersymmetries.

Our main interest lies in the case where $\gamma\rightarrow\pm\infty$, i.e., $\vert E\vert\rightarrow 1$. In this limit, one can naively expect that $\tilde{\epsilon}=\exp{(-\frac{\gamma}{2}\Gamma^{01})}\,\,\epsilon$ could vanish for the spinor components $\epsilon$ satisfying $\Gamma^{01}\,\epsilon=\pm\,\epsilon$, therefore equating both sides of Eq. (\ref{eigenspinor}). In the next section, we will perform explicit analysis on this.

\section{What if $E\rightarrow1$?}\label{sec4}
 
In the {\bf s}-basis of Ref. \cite{polchinski}, the eigenspinor $\epsilon$ of the operator (\ref{operator}) can be written as
\begin{eqnarray}\label{abcd}
\epsilon&=&(a, b, c, d)\nonumber\\
&\equiv& a\,\, \vert +1,\, +1,\, 2s_2, 2s_3, 2s_4>+b\,\, \vert +1,\, -1,\, 2s_2, 2s_3, 2s_4>\nonumber\\
&+&c\,\, \vert -1,\, +1,\, 2s_2, 2s_3, 2s_4>+d\,\, \vert -1,\, -1,\, 2s_2, 2s_3, 2s_4>.
\end{eqnarray}
The first and second entry $\pm1$'s denote the eigenvalues of $2S_0\equiv\Gamma^0\Gamma^9$ and $2S_1$ respectively. Since these two operators do not commute with the operator (\ref{operator}), the eigenspinor $\epsilon$ must be the sum of all possible eigenstates of $2S_0$ and $2S_1$. Due to the Weyl condition of $\epsilon$, the eigenvalues $s_a, (a=0,\cdots 4)$ are constrained by $\prod\limits_{a=0}^4 2s_a=1$. With $\lambda\equiv\exp{(-2i\sum\limits_{j=2}^k \phi_j s_j)}$, the eigenspinor equation reduces to
\begin{eqnarray}\label{eigencomp}
\left(\lambda\cos{\phi_1}-1\right)\left(a, b, c, d\right)-i\lambda\sin{\phi_1}\left[\cosh{\gamma}\left(a, -b, c, -d\right)-\sinh{\gamma}\left(d, -c, b, -a\right)\right]=0.
\end{eqnarray} 
In order for the above equation to have nontrivial solutions, $\exp{(-2i\sum\limits_{j=1}^k \phi_j s_j)}$ should be 1, which is just the condition used to analyse the cases with the magnetic fields only \cite{witten}. 

Let us look at the equation in the limit, $\gamma\rightarrow\infty$. There are some values of $\phi_1$ to consider first. At the exact points $\vert\phi_1\vert= 0, \pi$, the second and third term of Eq. (\ref{eigencomp}) vanish however large the parameter $\gamma$ becomes. The eigenspinor equation becomes independent of the boost then. The condition $\lambda=\pm1$ determines the number of the unbroken supersymmetries to be $2^{6-k}$ for the D0-D$(2k)$ brane systems ($k\ge 2$) with generic values of $\phi_{2,\cdots,k}$. In the D0-D2 brane system, $\lambda=1$, so 16 supersymmetries are preserved at $\phi_1=0$, while no supersymmetry is preserved at $\vert\phi_1\vert=\pi$. 

As for the values of $\phi_1$ other than $0, \pm\pi$, we keep the magnetic field
\begin{eqnarray}\label{magnetic}
B_1=-\frac{1}{\cosh{\gamma}}\cot{\phi_1}
\end{eqnarray}
to be finite in the limit $\gamma\rightarrow\infty$. The insertion of
\begin{eqnarray}\label{eq1}
\cos{\phi_1}&=&\mp\frac{B_1\cosh{\gamma}}{\sqrt{1+B_1^2\cosh^2{\gamma}}},\qquad \sin{\phi_1}=\pm\frac{1}{\sqrt{1+B_1^2\cosh^2{\gamma}}},
\end{eqnarray}
into Eq. (\ref{eigencomp}) yields
\begin{eqnarray}\label{eq2}
\left(\lambda B_1\pm\vert B_1\vert\right)\left(a,\,b,\,c,\,d\right)+i\lambda\left(a,\,-b,\,c,\,-d\right)-i\lambda\left(d,\,-c,\,b,\,-a\right)=0.
\end{eqnarray}
In the above two Eq.'s, the upper (lower) signs are for $\phi_1>0$ ($\phi_1<0$).
Only when $\lambda B_1\pm\vert B_1\vert=0$, the characteristic equation is satisfied and nontrivial solutions ($a=d$, $b=c$) are possible. Especially when $\vert\phi_1\vert=\frac{\pi}{2}$ (and $\gamma\rightarrow\infty$), the magnetic field $B_1$ vanishes and the condition $\lambda B_1\pm\vert B_1\vert=0$ is satisfied automatically without giving any constraint on $\lambda$. Therefore, 8 supersymmetries are preserved regardless of the values of other fields $B_{2,3,4}$. Let us explore the unbroken supersymmetries case by case for the rest of the values $\phi_1$.

\subsection{D0-D2 brane system}

With $\lambda=1$, the factor $(B_1\pm\vert B_1\vert)$ vanishes for $0<\vert\phi_1\vert<\frac{\pi}{2}$. Therefore 8 supersymmetries are preserved in these cases. As for $\vert\phi_1\vert>\frac{\pi}{2}$, no supersymmetry is preserved. Indeed in these latter cases, defining new angles $\vert\phi_1'\vert\equiv\vert\phi_1\mp\pi\vert<\frac{\pi}{2}$ results in extra minus sign in front of the operator (\ref{operator}), which implies that the system is then actually D0-$\overline{\mbox{D}2}$ brane system \footnote{We thank Joseph Polchinski for pointing this out to us.}. 

\subsection{D0-D4 brane system}

Different from the D0-D2 brane system, $\lambda=\exp{(-2i\phi_2s_2})$ can be either $+1$ or $-1$. As for $0<\vert\phi_1\vert<\frac{\pi}{2}$, the angle $\vert\phi_2\vert$ should vanish. On the other hand when $\vert\phi_1\vert>\frac{\pi}{2}$, the angle $\vert\phi_2\vert=\pi$. In both cases, $\vert B_2\vert\rightarrow\infty$ and the value $2s_2$ can be arbitrary, therefore 8 supersymmetries are preserved.

\subsection{D0-D6 brane system}

The equation $\lambda B_1\pm\vert B_1\vert=0$ is satisfied if $\lambda=\exp{(-2i\phi_2s_2-2i\phi_3s_3)}=1$ for $0<\vert\phi_1\vert<\frac{\pi}{2}$, and $\lambda=-1$ for $\vert\phi_1\vert>\frac{\pi}{2}$. In the former case, let $\phi_2+\phi_3=0$. For generic values of $\phi_2=-\phi_3$, the values of $s_{2,3}$ are constrained as $s_2=s_3$. On the other hand for $\phi_2=-\phi_3=0$, the values $s_{2,3}$ are free. Therefore, 4 supersymmetries are preserved for generic values of $B_{2}=-B_{3}$ while 8 supersymmetries are preserved when $\vert B_{2}=-B_{3}\vert\rightarrow\infty$. The same conclusion can be derived for the latter case, that is, $\vert\phi_1\vert>\frac{\pi}{2}$.

\subsection{D0-D8 brane system}

In order to preserve any possible supersymmetry, $\lambda=\exp{(-2i\phi_2s_2-2i\phi_3s_3-2i\phi_4s_4)}=1$ for $0<\vert\phi_1\vert<\frac{\pi}{2}$, and $\lambda=-1$ for $\vert\phi_1\vert>\frac{\pi}{2}$. In the former case, let $\phi_2+\phi_3+\phi_4=0$. For the generic values of $\phi_{2,3,4}$, the values of $s_{2,3,4}$ are constrained as $s_2=s_3=s_4$. Due to the Weyl condition, the eigenspinor must be of the form,
\begin{eqnarray}
\epsilon=a\left[\left(+++++\right)+\left(--+++\right)\right]+b\left[\left(+----\right)+\left(-+---\right)\right].
\end{eqnarray}
Therefore 2 supersymmetries are preserved for the generic values of $B_{2,3,4}$ satisfying $B_2B_3+B_3B_4+B_4B_2=1$. If $\phi_2=0$ and $\phi_{3,4}$ are generic, the value $s_2$ can be arbitrary and $s_3=s_4$. Due to the Weyl condition, the eigenspinor is of the form 
\begin{eqnarray}
\epsilon&=&a\left[\left(+++,\,2s_3,\,2s_3\right)+\left(--+,\,2s_3,\,2s_3\right)\right]\nonumber\\
&&+b\left[\left(+--,\,2s_3,\,2s_3\right)+\left(-+-,\,2s_3,\,2s_3\right)\right],
\end{eqnarray}
which shows us that 4 supersymmetries are preserved when $\vert B_2\vert\rightarrow\infty$ and $B_3=-B_4$ takes a generic value. Especially when $\phi_2=\phi_3=\phi_4=0$, the values of $s_{2,3,4}$ can be arbitrary but subject to the Weyl condition. In other words, 8 supersymmetries are preserved when $\vert B_{2,3,4}\vert\rightarrow\infty$. One can easily draw the same conclusion for the case when $\vert\phi_1\vert>\frac{\pi}{2}$. We summarize the results in the following table.
\begin{table}[h]
\begin{tabular}{||l|c||c|c|c||}
\hline
system& $\vert B_{2,3,4}\vert$ & $0<\vert\phi_1\vert<\frac{\pi}{2}$ & $\vert\phi_1\vert=\frac{\pi}{2}$ & $\frac{\pi}{2}<\vert\phi_1\vert$\\ \hline\hline
D0-D2 & $\cdot$ & 8 &  8 & 0 \\ \hline
D0-D4 & $\vert B_2\vert\rightarrow\infty$ & 8 & 8 & 8 \\ \cline{2-5}
& $\vert B_2\vert<\infty$ & 0 & 8 & 0 \\ \hline
D0-D6 & $\vert B_2\vert=\vert B_3\vert\rightarrow\infty$  & 8 & 8 & 8 \\ \cline{2-5}
  & $\vert B_2\vert=\vert B_3\vert<\infty$  & 4 & 8  & 4 \\ \cline{2-5}
 & $\vert B_2\vert\ne\vert B_3\vert$ &  0 & 8 & 0 \\ \hline
D0-D8 & $\vert B_{2,3,4}\vert\rightarrow\infty$ & 8 & 8 & 8 \\ \cline{2-5}
 & $\vert B_2\vert\rightarrow\infty$, $\vert B_3\vert=\vert B_4\vert<\infty$ & 4 &  8 & 4 \\ \cline{2-5}
 &  $B_2B_3\pm B_3B_4\pm B_4B_2=1$ & 2 & 8 & 2 \\ \cline{2-5}
  & otherwise & 0 & 8 & 0   \\ \hline
\end{tabular}
\caption{\footnotesize Supersymmetries preserved by D0-D$(2k)$ brane systems when $E\rightarrow 1$ keeping $B_1=-\cot{\phi_1}\sqrt{1-E^2}$ finite.} 
\end{table}

\section{Intersecting D-strings under a uniform Lorentz boost}\label{sec5}

In this section, we explore the supersymmetries preserved by two intersecting D-strings in uniform motion. We consider the situation where those two D-strings are boosted together so that they are not in relative motion. The supersymmetries preserved by the first and the second D-string are 
\begin{eqnarray}
&&Q+\bar{\beta}^2\beta_2^\bot\tilde{Q}, \qquad Q+\bar{\beta}'^2\beta_2^\bot\tilde{Q},\nonumber\\
&&\qquad \bar{\beta}^2=\rho(\gamma)\rho(\phi)\beta^2\rho(-\phi)\rho(-\gamma),\quad
\bar{\beta}'^2=\rho(\gamma)\rho(\phi')\beta^2\rho(-\phi')\rho(-\gamma),
\end{eqnarray} 
respectively. Therefore they are moving together with the velocity, $v\equiv\tanh{\gamma}$ and intersecting at an angle, $\vert\phi'-\phi\vert$ (hereafter the intersecting angle is defined as the difference between tilting angles, namely $\vert\phi'-\phi\vert$ here). The supersymmetries preserved by both D-strings are determined by the spinors invariant under 
\begin{eqnarray}
\beta_2^{-\bot}(\bar{\beta}^2)^{-1}\bar{\beta}'^2\beta_2^\bot=\rho(\gamma)\rho(-2(\phi'-\phi))\rho(-\gamma).
\end{eqnarray}
This operator is the same as that of D0-D2 brane system in Eq. (\ref{operator}) except that $\phi_1$ is replaced by $\phi'-\phi$.
 Therefore for the velocity, $v<1$, 16 supersymmetries are preserved when the intersecting angle vanishes. Otherwise, no supersymmetry is preserved. 

In the limit $v\rightarrow 1$, the supersymmetric loci (in the parameter space) become different in general. From the results of the previous section, we easily note that the vanishing intersecting angle $\vert\phi'-\phi\vert=0$ still ensures 16 supersymmetries. This implies that 16-supersymmetric locus is just the same as that of the finite boost case ($v<1$). However, there are other `less supersymmetric' loci in the infinite Lorentz boost limit. Let us start with the case where one of $\phi$, $\phi'$ is zero, say, $\phi=0$ for simplicity. Again from the results of the previous section, we note that 8 supersymmetries are preserved when $0<\vert\phi'\vert\le\frac{\pi}{2}$. 

From here on, let us assume $\vert\phi'-\phi\vert\ne0$ and $\vert\phi',\phi\vert\ne0,\pi$ so that $B'$ and $B$ be finite. Then the angles $\phi'$ and $\phi$ are related to $B'$ and $B$ as in (\ref{eq1}). Especially when $\phi=\pm\frac{\pi}{2}$, the magnetic field $B=0$ exactly and the eigenspinor equations become those in Eq. (\ref{eigencomp}) with $\lambda=1$ and $\phi_1$ replaced by $\phi'\mp\frac{\pi}{2}$. Therefore, we obtain
\begin{eqnarray}
a\left(1\mp\vert B'\vert\cosh{\gamma}-iB'\cosh^2{\gamma}\right)+idB'\cosh{\gamma}\sinh{\gamma}=0,\nonumber\\
iaB'\cosh{\gamma}\sinh{\gamma}-d\left(1\mp\vert B'\vert\cosh{\gamma}+iB'\cosh^2{\gamma}\right)=0,\nonumber\\
b\left(1\mp\vert B'\vert\cosh{\gamma}+iB'\cosh^2{\gamma}\right)-icB'\cosh{\gamma}\sinh{\gamma}=0,\nonumber\\
ibB'\cosh{\gamma}\sinh{\gamma}+c\left(1\mp\vert B'\vert\cosh{\gamma}-iB'\cosh^2{\gamma}\right)=0.
\end{eqnarray}
In the $\gamma\rightarrow\infty$ ($v\rightarrow\infty$) limit, $a= d$ and $b= c$ solve the above equations if $B'\ne 0$, therefore 8 supersymmetries are preserved then. 

Now we consider the other cases where neither of $\vert\phi'\vert$ and $\vert\phi\vert$ is $\frac{\pi}{2}$, so both $B'$ and $B$ are nonvanishing. For simplicity, let us set $0<\phi'<\frac{\pi}{2}$ so that $B'<0$ and classify the system into four classes according to the range of $\phi$. In terms of finite values $B'$ and $B$,
\begin{eqnarray}
\cos{(\phi'-\phi)}&=&\pm\frac{1+B'B \cosh^2{\gamma}}{\sqrt{\left(1+B'^2\cosh^2{\gamma}\right)\left(1+B^2\cosh^2{\gamma}\right)}},\nonumber\\
\sin{(\phi'-\phi)}&=&\pm\frac{\left(B'-B\right) \cosh{\gamma}}{\sqrt{\left(1+B'^2\cosh^2{\gamma}\right)\left(1+B^2\cosh^2{\gamma}\right)}},
\end{eqnarray}
where the upper(lower) sign is for the range $\phi>0$  ($\phi<0$).
In the $\gamma\rightarrow\infty$ limit, the eigenspinor equation (\ref{eigencomp}), with appropriate substitution of $\lambda=1$ and $\phi_1\rightarrow\phi'-\phi$, becomes satisfied when
\begin{eqnarray}\label{condition}
\cos{(\phi'-\phi)}\approx\mbox{sgn}(B')\,\,\mbox{sgn}(B)=\pm1.
\end{eqnarray}
The eigenspinor components are then constrained as $a=d$, $b=c$ ensuring 8 supersymmetries. There are two classes which meet the above condition (\ref{condition}). These are the cases i) $0<\phi'<\frac{\pi}{2},\,$ $0<\phi<\frac{\pi}{2}$ and ii) $0<\phi'<\frac{\pi}{2},\,$ $-\frac{\pi}{2}<\phi<0$, where 8 supersymmetries are preserved. In the other two cases, no supersymmetry is preserved. 

These results look confusing because the latter cases that do not preserve any supersymmetry can be obtained just by rotating one of D-strings by an angle $\pm\pi$. However here again, this $\pi$-rotation corresponds to the orientation flip. 

The situations can be viewed in the T-dual set-up as follows. Let us take a T-duality transformation, on the whole system, along the boost direction. Each tilted D-string becomes a D2-brane (or a $\overline{\mbox{D}2}$-brane depending on the orientation) with the magnetic fields $B'$ and $B$. The momenta carried by the D-strings are T-dualized to the electric fields along the boost direction over the world volumes of the (anti-)D2-branes.

Fig. 1 and Fig. 2 summarize the results. In Fig. 1, the upper right figure shows two D2-branes with the BI electric field $E=1$ and parallel magnetic fields $B'$ and $B$ on each D2-brane respectively. When $B'=B$, 16 supersymmetries are preserved. This is valid for arbitrary value of $E$. Therefore this 16-supersymmetric locus is continuously connected to the familiar configuration of two D2-branes without any world volume field. At the point $E=1$, the difference $B'-B$ triggers the supersymmetry breaking by half, so leaves 8 supersymmetries unbroken. For the noncritical value of $E<1$, no supersymmetry is preserved if $B'-B\ne0$. The lower right figure of Fig. 1 shows just the configuration that drew recent interests \cite{bak,mateos,ohta}. This cannot be connected continuously to the well-known nonsupersymmetric configuration of a D2-$\overline{\mbox{D}2}$-brane pair. Indeed, the magnetic fields $B'$ and $B$ can be turned off only by an infinite boost (see Eq. (\ref{magnetic})). That is to say, the angles $\vert\phi'\vert$ and $\vert\phi\vert$ cannot be tuned to the value $\pi/2$ due to the condition (\ref{condition}). Therefore turning off the magnetic fields inevitably increases the electric field $E$ to 1. Fig. 2 shows two nonsupersymmetric configurations. To sum up, two intersecting D-strings in uniform motion preserve 16 supersymmetries when the intersecting angle vanishes, 8 supersymmetries when $\vert\phi'\ne\phi\vert<\pi/2$ or $\vert\phi'\ne\phi\vert>\pi/2$, and no supersymmetry otherwise.

\epsfbox{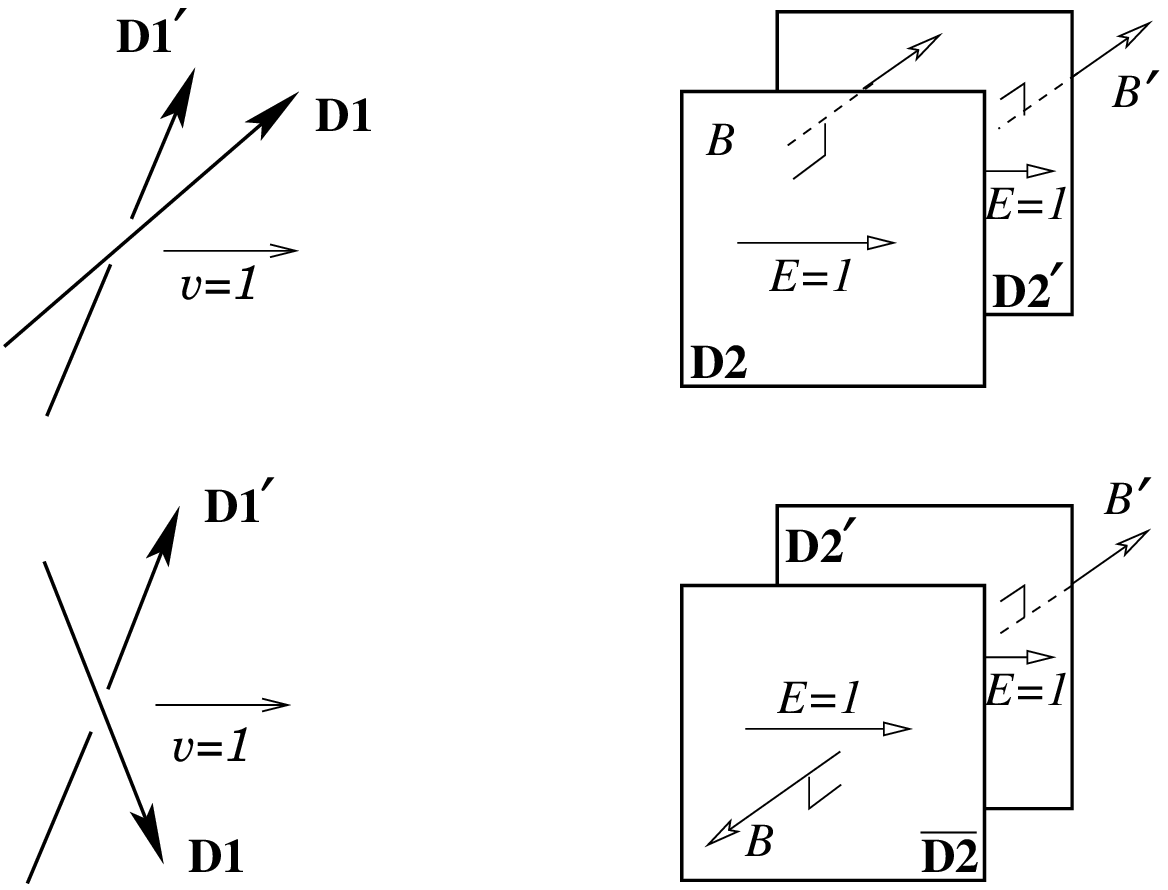}
\begin{flushleft}
{\footnotesize Fig. 1: Two D-strings are in uniform motion at the speed of light (Left). Depending on their orientations, T-dual transformation along their moving direction generates a D2-D2 or a D2-$\overline{\mbox{D}2}$ pair (Right). The tilting angles of the D-strings determine the signature of the magnetic fields in the T-dual picture. All these configurations preserve 8 supersymmetries.
}
\end{flushleft}
\epsfbox{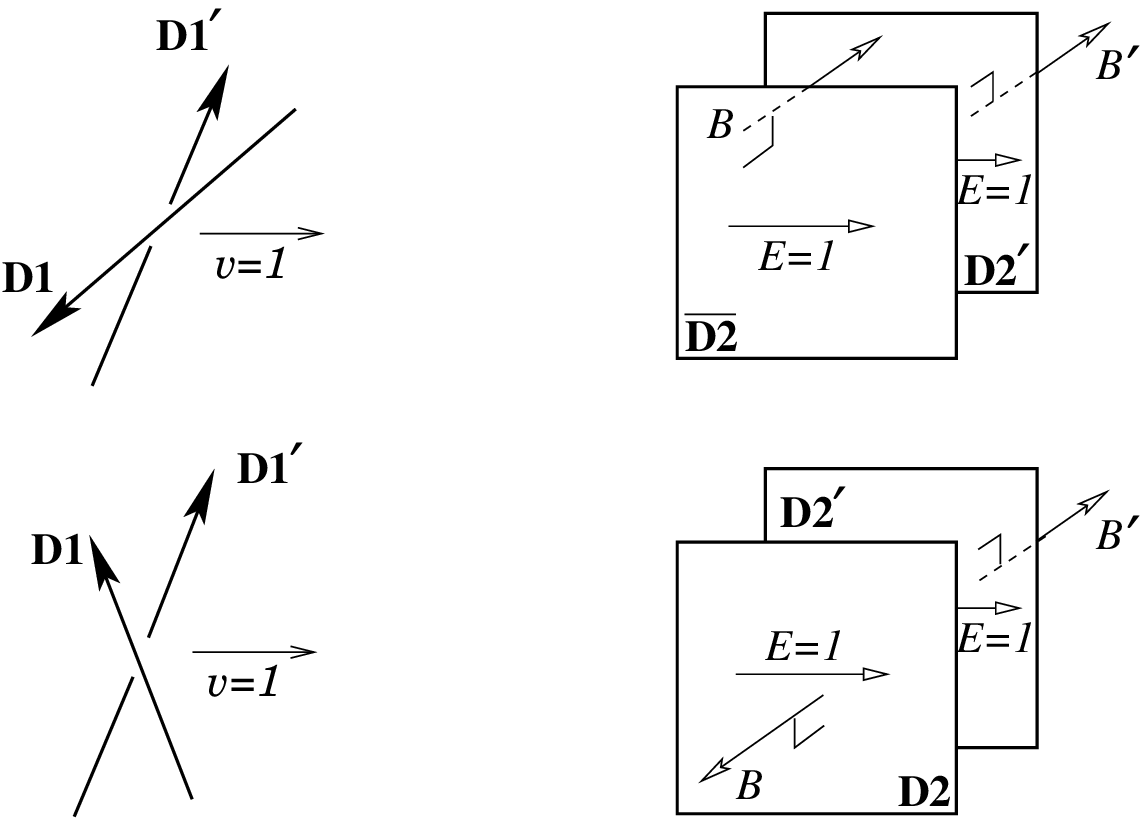}
\begin{flushleft}
{\footnotesize Fig. 2: D$1'$ is oriented forward while D1 is oriented backward with respect to the moving direction. No supersymmetry is preserved for these configurations.
}
\end{flushleft}

\section{D-strings in relative motion}\label{sec6}

In this section, we allow relative motion between two D-strings and explore the preserved supersymmetries. We restrict our attention to only the cases where the mutual distance between D-strings is kept fixed. Other cases could break supersymmetries by the velocity dependent force as in the D0-D0 scattering \cite{becker}. The simplest case can be composed as follows. Let the first D-string be static but tilted by an angle $\phi$ with respect to the $x_1$-axis. The supersymmetry preserved by this D-string is given by
\begin{eqnarray}
Q+\rho(\phi)\beta^2\rho(-\phi)\beta_2^{\bot}\tilde{Q}.
\end{eqnarray}
Let the second D-string lie along $x_2$-axis and boosted along $x_1$-direction with the boost parameter $\gamma$. The supersymmetry preserved by this second D-string is
\begin{eqnarray}\label{2ndstring}
Q+\rho(\gamma)\beta^1\rho(-\gamma)\beta_2^{\bot}\tilde{Q}.
\end{eqnarray}
The unbroken supersymmetries preserved by both D-strings are determined by the spinors invariant under
\begin{eqnarray}
&&\beta_2^{-\bot}\rho(\gamma)\left(-\beta^1\right)\rho(-\gamma)\rho(\phi)\beta^2\rho(-\phi)\beta_2^{\bot}\nonumber\\
&&=\exp{\left[\gamma\Gamma^{01}\right]}\exp{\left[\tilde{\phi}\Gamma^{12}\right]}\nonumber\\
&&=\exp{\left[\tilde{\phi}\Gamma^{12}\right]}\cosh{\gamma}+\exp{\left[-\tilde{\phi}\Gamma^{12}\right]}\Gamma^{01}\,\sinh{\gamma},
\end{eqnarray}
where $\tilde{\phi}=\frac{\pi}{2}-\phi$ and its magnitude $\vert\tilde{\phi}\vert$ corresponds to the intersecting angle. 

Although this is the simplest case which allows relative motion between the D-strings, it actually exhausts all the cases which can be reached by finite Lorentz boosts on the whole system. Supersymmetry is intact about these extra boost, $\rho(\bar{\gamma})=\exp{(i\bar{\gamma}\Sigma^{0i})}$, because the above operator is changed only by a similar transformation;
\begin{eqnarray}
&&\beta_2^{-\bot}\rho(\bar{\gamma})\rho(\gamma)\left(-\beta^1\right)\rho(-\gamma)\rho(\phi)\beta^2\rho(-\phi)\rho(-\bar{\gamma})\beta_2^{\bot}\nonumber\\
&&=\left\{
\begin{array}{ll}
\rho(-\bar{\gamma})\exp{\left[\gamma\Gamma^{01}\right]}\exp{\left[\tilde{\phi}\Gamma^{12}\right]}\rho(\bar{\gamma})& \quad\mbox{(boost in the $(x_1, x_2)$-plane)},\\
\rho(\bar{\gamma})\exp{\left[\gamma\Gamma^{01}\right]}\exp{\left[\tilde{\phi}\Gamma^{12}\right]}\rho(-\bar{\gamma})& \quad\mbox{(boost transverse to the $(x_1, x_2)$-plane)}.\nonumber\\
\end{array}
\right.
\end{eqnarray}

A pure longitudinal motion (along $x_2$-direction) of the second D-string does not affect the supersymmetry as we see easily in Eq. (\ref{2ndstring}). The situation becomes exactly the same as the case of two static D-strings intersecting at an angle $\vert\tilde{\phi}\vert$. One could imagine more general motion of the second string along $(x_1, x_2)$-plane. Without loss of generality, this can be achieved by a combination of tilting and boosting the second D-string. However, this complicated situation does not give us any simple insight, so we will not consider the case.

In the representation (\ref{abcd}), the eigenspinor equation becomes
\begin{eqnarray}\label{eigencomp2}
&&\left(\cosh{\gamma}\cos{\tilde{\phi}}-1\right)\left(a,\,b,\,c,\,d\right)+i\cosh{\gamma}\sin{\tilde{\phi}}\left(a,\,-b,\,c,\,-d\right)\nonumber\\
&&\quad +\sinh{\gamma}\cos{\tilde{\phi}}\left(d,\,c,\,b,\,a\right)-i\sinh{\gamma}\sin{\tilde{\phi}}\left(d,\,-c,\,b,\,-a\right)=0.
\end{eqnarray} 
Suppose $\gamma<\infty$. In order for the above equation to have any nontrivial solution, its characteristic equation should be satisfied, namely,
\begin{eqnarray}\label{lc}
\cosh{\gamma}\cos{\tilde{\phi}}=1.
\end{eqnarray}
The insertion of this condition into Eq. (\ref{eigencomp2}) yields
\begin{eqnarray}
&&i\,a\tan{\tilde{\phi}}\pm d\tan{\tilde{\phi}}\left(\cos{\tilde{\phi}}-i\sin{\tilde{\phi}}\right)=0,\nonumber\\
&&i\,b\tan{\tilde{\phi}}\mp c\tan{\tilde{\phi}}\left(\cos{\tilde{\phi}}+i\sin{\tilde{\phi}}\right)=0,
\end{eqnarray}
Note that the condition (\ref{lc}) implies that $\vert\tilde{\phi}\vert<\pi/2$. Especially when $\tilde{\phi}=0$, the above eigenspinor equations are satisfied automatically. This is precisely when $\cosh{\gamma}=\cos^{-1}{\tilde{\phi}}=1$, namely, the moving D-string comes to rest and the intersecting angle vanishes; the familiar configuration of two parallel D-strings preserving 16 supersymmetries. For the values $0<\vert\tilde{\phi}\vert<\pi/2$, the above result shows that 8 supersymmetries are preserved.

Let us see the physical meaning of the condition (\ref{lc}). In terms of the velocity $v=\tanh{\gamma}$ of the moving D-string, $\cosh{\gamma}=1/\sqrt{(1-v^2)}$. The condition tells us that the intersecting point of the two D-strings should move at the speed of light, 
\begin{eqnarray}\label{supercond}
\tilde{v}=\frac{v}{\sin{\tilde{\phi}}}=1.
\end{eqnarray}
The same configuration (named as `null scissors') was recently discussed in \cite{bachas,myers}. Its supersymmetry was checked in a different manner: Various duality chains lead the null scissors to the null intersecting M5-branes. This latter configuration is known to preserve 8-supersymmetries \cite{acharya}.

It is remarkable again that Eq. (\ref{lc}) implies $\vert\tilde{\phi}\vert<\pi/2$. If we invert the static D-string, thereby make the intersecting angle larger than $\pi/2$, the configuration does not preserve any supersymmetry even though the intersecting point of the D-strings moves at the speed of light. Widely open null scissors are not supersymmetric.

Like other configurations discussed in the previous sections, there is a special limit ($\gamma\rightarrow\infty$), where the supersymmetries are ensured. The eigenspinor equation (\ref{eigencomp2}) is satisfied when $\exp{(i\tilde{\phi})}a+\exp{(-i\tilde{\phi})}d=0$ and $\exp{(-i\tilde{\phi})}b+\exp{(i\tilde{\phi})}c=0$. This implies that 8 supersymmetries are always preserved in the limit $\gamma\rightarrow\infty$, regardless of the value of $\tilde{\phi}$. This case allows the angle out of the range $\vert\tilde{\phi}\vert<\pi/2$. The speed of the intersecting point is $\tilde{v}=\csc{\tilde{\phi}}\ge 1$, therefore superluminal in general. Only when $\tilde{\phi}\rightarrow\frac{\pi}{2}$, its speed approaches the speed of light. 

\epsfbox{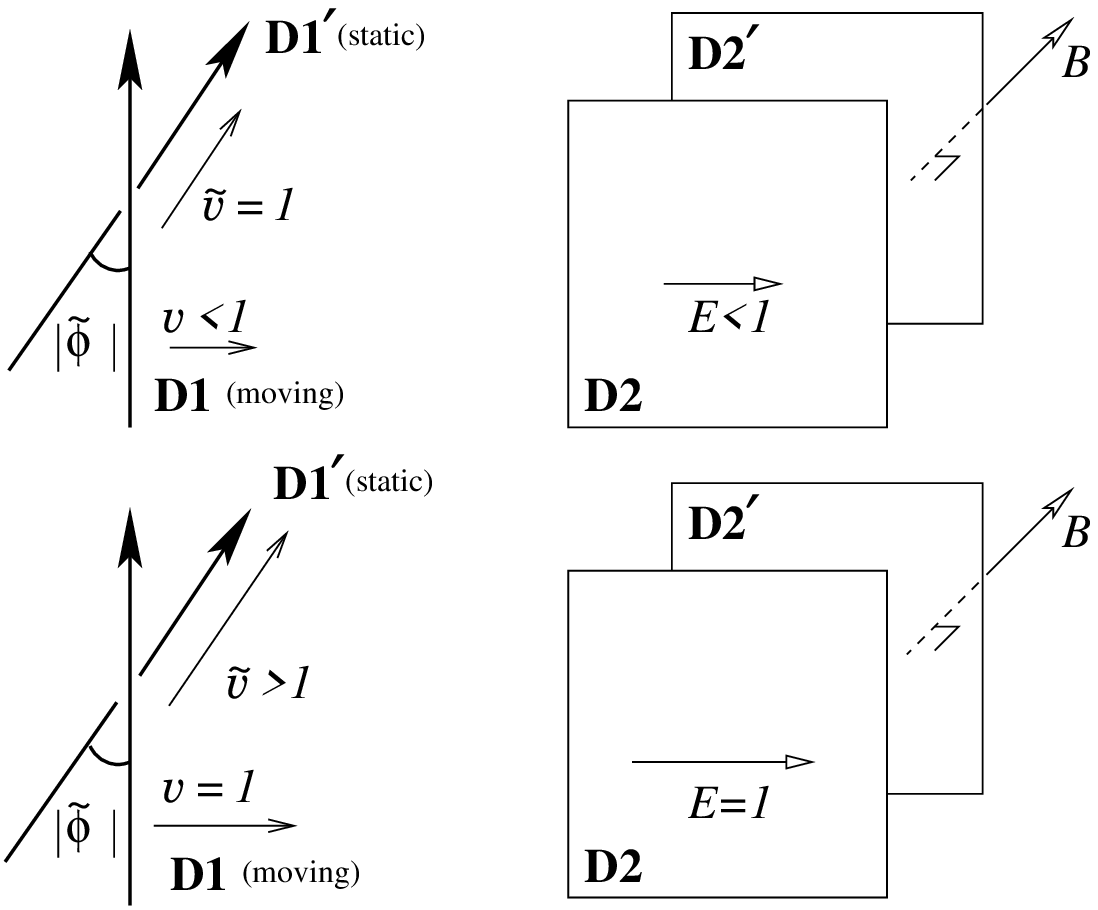}
\begin{flushleft}
{\footnotesize Fig. 3: Two D-strings are in relative motion (D$1'$: static, and D1: moving). When the motion is timelike, 8 supersymmetries are preserved when the intersecting angle is less than $\pi/2$ and the intersection point moves at the speed of light. If the motion is lightlike, the intersection point will move with the superluminal speed. 8 supersymmetries are preserved regardless of the value of the intersecting angle.
}
\end{flushleft}

Let us examine the supersymmetric loci in the parameter space composed of $\tilde{\phi}$ and $v$.
The upper left figure of Fig. 3 shows the `null scissors' configuration discussed in \cite{bachas,myers}. One interesting limit of this configuration is the closed null scissors, namely the limit where the intersecting angle vanishes. The supersymmetric condition (\ref{lc}) then regulates the moving D-string so that it come to rest. Our analysis shows that just at this end point of 8-supersymmetric locus, (determined by Eq. (\ref{supercond})), the 16-supersymmetric locus begins. See Fig. 4 which shows the whole supersymmetric loci. At the other end point of the locus of the null scissors, where the intersecting angle approaches $\pi/2$, a new 8-supersymmetric locus begins: The moving D-strings moves at the speed of light at the point. The intersecting angle can be arbitrary as long as the motion of the moving D-string is lightlike. (See the lower left figures of Fig. 3 and Fig. 5.)

Along with this new locus, we meet an interesting point where the intersecting angle vanishes. The case describes two parallel D-strings, one of which is under an infinite boost. Although their mutual distance varies significantly with time, the configuration preserves 8 supersymmetries. This result is very puzzling because two parallel D-branes in relative motion interact with each other via velocity dependent force. However, one could naively understand the situation as follows. Each of two D-strings is precisely at the lightcone of the other D-string, namely on the horizon. Therefore their mutual interaction cannot occur. The same understanding could be made for the case where they are anti-parallel ($\vert\tilde{\phi}\vert=\pi$). Further studies on these are in need. The upper left figure of Fig. 5 shows a null scissors configuration but no supersymmetry is preserved because the intersecting angle is larger than $\pi/2$.

\epsfbox{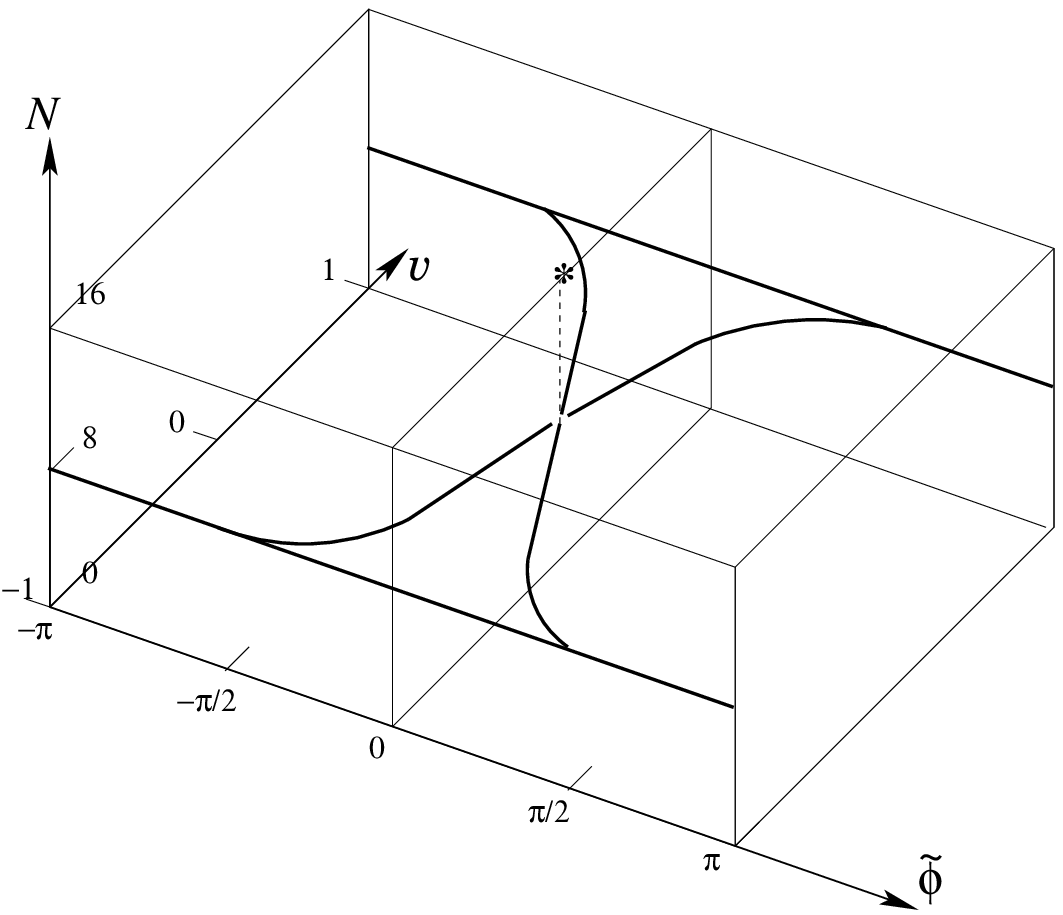}
\begin{flushleft}
{\footnotesize Fig. 4: The bold lines show 16- and 8-supersymmetric loci of two D-strings in relative motion. $\tilde{\phi}$ represents the intersecting angle (ranging from $-\pi$ to $\pi$) and $v$ represents the speed of the moving D-string (ranging from $-1$ to $1$). $N$ denotes the number of the preserved supersymmetry. The curves $v=\pm\sin{\tilde{\phi}}$ drawn for $-\pi/2<\tilde{\phi}<\pi/2$ show the supersymmetric loci of the null scissors configuration. They join with those (two thick horizontal lines) of the configuration where the motion of the moving D-string is lightlike. The loci jump at the origin to 16-supersymmetric point ($\ast$) representing two static parallel D-strings.}
\end{flushleft}

Under the T-dual transformation along the moving direction, the static D-string become a D2- or a $\overline{\mbox{D}2}$-brane depending on its orientation. See the right figures of Fig.3 and Fig. 5. The tilting angle makes the magnetic field $B$. The moving D-string becomes a D2-brane with the electric field $E$ on its world volume. The supersymmetric condition (\ref{supercond}) for the case $E<1$ becomes
\begin{eqnarray}\label{eb}
E^2+E^2B^2-B^2=0.
\end{eqnarray} 
\\

\epsfbox{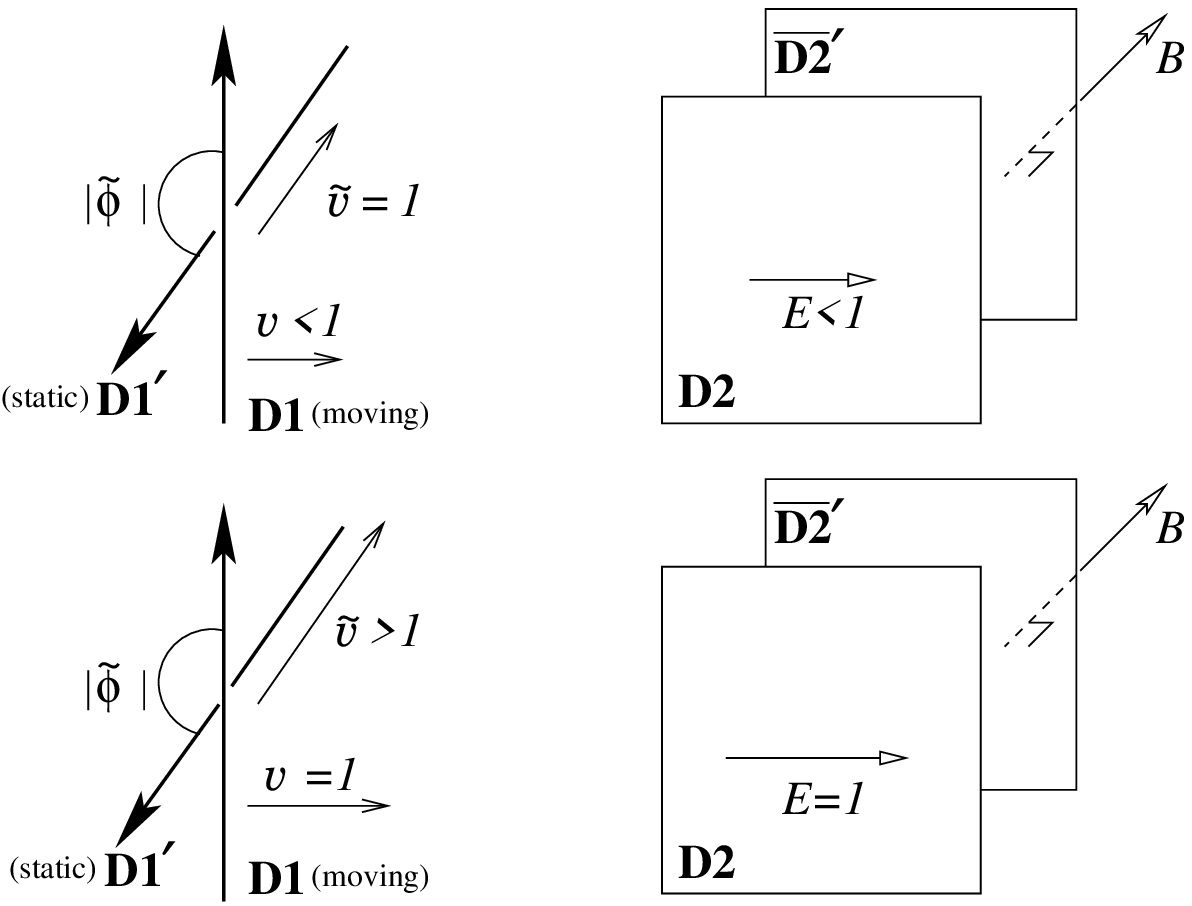}
\begin{flushleft}
{\footnotesize Fig. 5: When the intersecting angle exceeds $\pi/2$, no supersymmetry is preserved even when the intersection point moves at the speed of light (Upper). However, if the string D1 moves at the speed of light, the intersecting angle does not matter and 8 supersymmetries are always preserved (Lower).
}
\end{flushleft}

\section{Discussions}\label{sec7}

In this paper, we considered various configurations of D0-D($2k$) brane systems and the intersecting D-strings which preserve some fractions of 32 supersymmetries. In the former systems, we especially focused on the role of the BI electric field on the preserved supersymmetries. In the latter systems, we check the supersymmetries when these two D-strings are in uniform or relative motion. Now, let us think over the situations in various dual pictures. 

\subsection{T-dual picture of D0-D($2k$) brane systems}

Under the T-duality along the electric field direction ($x_1$-axis), the system is dualized to a D1-D$(2k-1)$ brane system, where the D-string extends along $x_1$-direction and the D$(2k-1)$ brane is tilted by an angle $\phi_1$ in $(x_1, x_2)$-plane and moving along $x_1$-direction. Although they are in relative motion, the direction of the motion is not transverse to the D$(2k-1)$ brane, thus, the situation does not overlap with the cases discussed in Sec. \ref{sec6}. As for D1-D1 case, one can draw the whole picture of the supersymmetric loci in the parameter space composed of $(\phi_1,\,v,\,N)$. In Fig. 6, $\phi_1$ is the tilting angle of the moving D-string, $v$ is its speed, and $N$ is the number of the preserved supersymmetries. 

\epsfbox{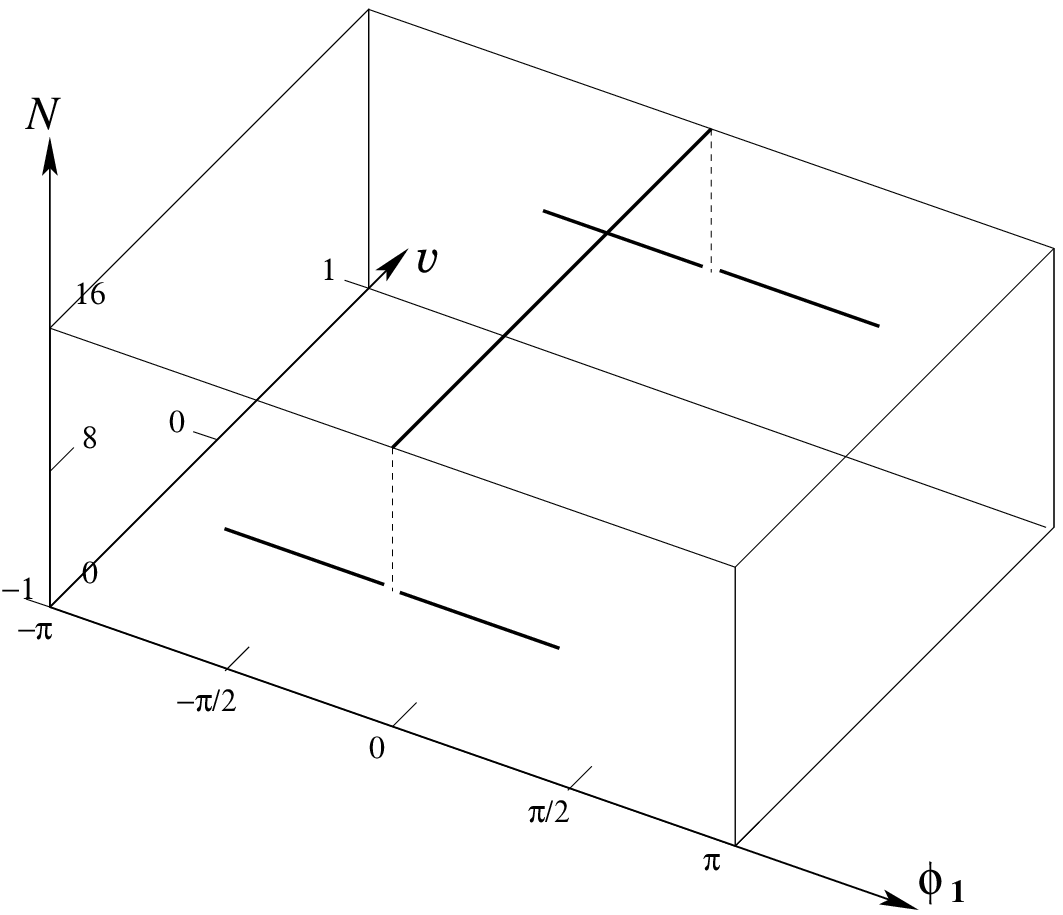}
\begin{flushleft}
{\footnotesize Fig. 6: The figure shows 16- and 8-supersymmetric loci of two intersecting D-strings under a uniform boost. One of the D-strings extends to the $x_1$-direction and $\phi_1$ represents the intersecting angle (ranging from $-\pi$ to $\pi$) and $v$ represents the speed of the moving D-string (ranging from $-1$ to $1$). $N$ denotes the number of the preserved supersymmetry. The 8-supersymmetric loci are broken at $\phi_1=0$, where they jump to the 16-supersymmetric locus. }
\end{flushleft}

When $\phi_1=0$, 16 supersymmetries are preserved regardless of the value of $v$. When $\phi_1\ne0$, no supersymmetry is preserved for generic values of $v$, but 8 supersymmetries are preserved if the motion of the moving D-string is lightlike. 

This `supersymmetry enhancement' at $v=1$ is actually a common feature shown in all the other cases also. It is due to a different way of satisfying the eigenspinor equation (\ref{eigencomp}) in the infinite boost limit: $\Gamma^{01}\epsilon=\epsilon$ (thus reducing the supersymmetries by half) provides new supersymmetric loci $\lambda B_1\pm\vert B_1\vert=0$ when $E=1$ (as we see in Eq. (\ref{eq2})).

It is interesting to see that similar behavior happens, in the name of the `supernumerary supersymmetry', in the Penrose limits of some `less supersymmetric' $AdS$ spaces \cite{cvetic,gauntlett}. (Penrose limit focusses on the geometry near the lightlike geodesic.) For example, in the Penrose limit of $AdS_3\times S^3\times T^4$ geometry \cite{cho2}, two gamma matrix components $\gamma^{(0)}$ and $\gamma^{(1)}$ in the global coordinates of $AdS$ geometry coalesce to a projector $\Gamma^{(+)}$ in the plane wave geometry, thereby make the dilatino Killing spinor equation factorized as $(\gamma^{(0)(2)(3)}+\gamma^{(1)(4)(5)})\epsilon\rightarrow\Gamma^{(+)}(\Gamma^{(2)(3)}+\Gamma^{(4)(5)})\epsilon=0$. Each of the factors can annihilate the spinor $\epsilon$ separately. Especially the second factor ensures 8 additional symmetries.

Supersymmetric loci for the other D1-D($2k-1$) brane systems ($k>1$) are difficult to draw because the parameter space is higher dimensional composed of $(\phi_1,\cdots\phi_k,\,v,\,N)$. Fig. 7 shows the supersymmetric loci for D1-D3 brane system exhibited in the space $(\phi_1,\,\phi_2,\,v)$ only. The bold vertical line at the center represents 16-supersymmetric locus, while other loci (two planes intersecting at right angle for the part $v<1$ and three intersecting lines at $v=1$) are of 8 supersymmetries. Generally in the D1-D($2k-1$) brane system ($k\ne1$), the 8-supersymmetric loci $\vert\phi_1\vert=\pi/2$ at $v=1$ touches those of $v<1$ part at some finite number of points.  

\epsfbox{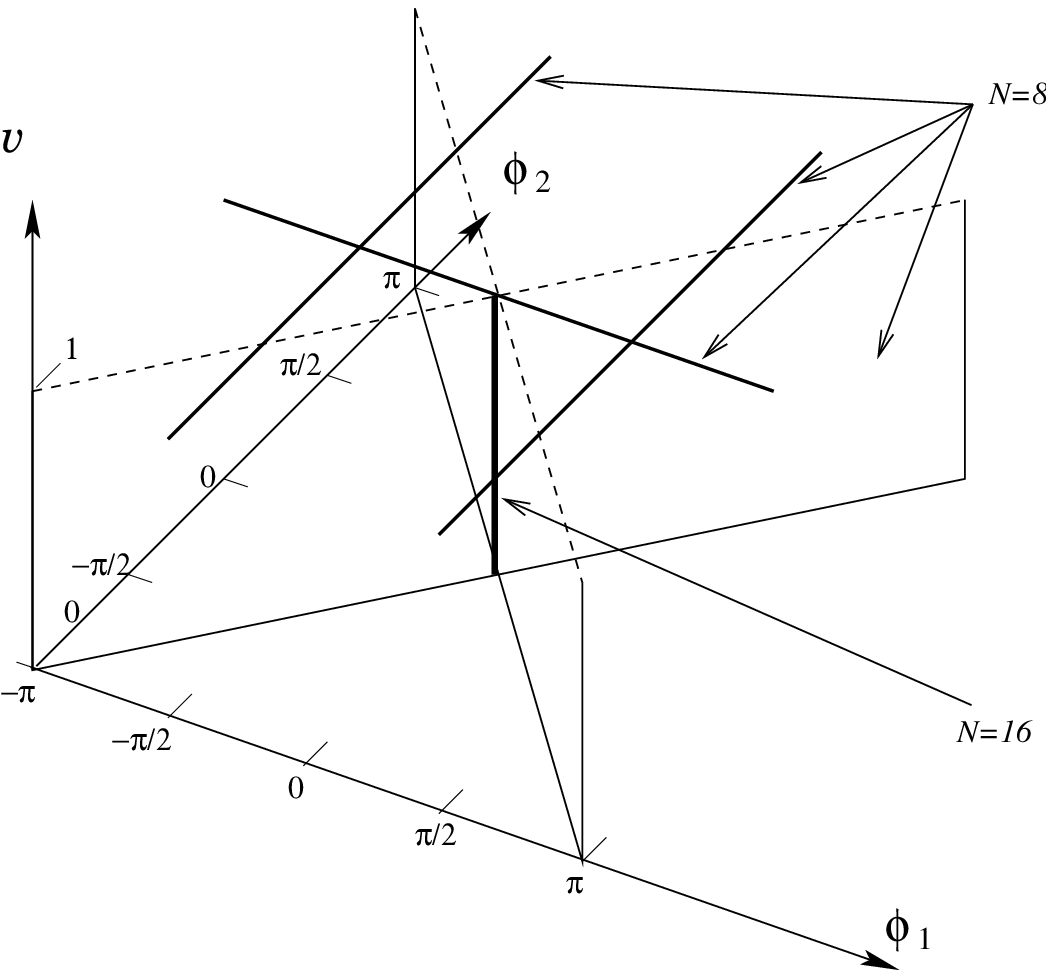}
\begin{flushleft}
{\footnotesize Fig. 7: The figure shows 16- and 8-supersymmetric loci of D1-D3 brane system. The D3 brane is tilted by an angle $\phi_1$ in $(x_1,x_2)$-plane and has BI magnetic field $B_2=-\cot{\phi_2}$ over $(x_3, x_4)$-plane. It is moving in the $x_1$-direction at the speed $v$. The D1 brane is static and extends along $x_1$-direction. The part $v<0$ is suppressed but is the same as the part shown here.}
\end{flushleft}

\subsection{M-brane configuration}

Let us consider the D2-D2 or D2-$\overline{\mbox{D}2}$ brane systems discussed in Sec. \ref{sec5} and \ref{sec6} in the S-dual set-up.  
Upon the lift to the 11 dimensions, a fundamental string becomes a longitudinal M2-branes (wrapping $x_{11}$-direction), a D0-brane becomes a M-wave quanta carrying a basic unit of the momentum (around $x_{11}$-direction), and a D2- or a $\overline{\mbox{D}2}$-brane becomes a M2- or a $\overline{\mbox{M}2}$-brane lying transversely to the $x_{11}$-direction. A D2-brane with BI fields, even though it is considered as a D2-brane with D0-branes or fundamental strings uniformly melted over its world volume, becomes a tilted M2-brane moving along the compact $x_{11}$-direction. This S-dual picture is very similar to the T-dual picture (a tilted D-string in motion). The geometrical configuration of the M2-brane is explicitly determined by the following mapping \cite{schmidhuber,alwis};
\begin{eqnarray}\label{lll}
\partial_t x_{11}&=&-\frac{B}{\sqrt{1-E^2+B^2}},\nonumber\\
\partial_1 x_{11}&=&0,\nonumber\\
\partial_2 x_{11}&=&\frac{E}{\sqrt{1-E^2+B^2}}.
\end{eqnarray}
We see, from Eqs. (\ref{angle}), (\ref{velocity}), and (\ref{bifield}), the geometric configuration in the T-dual picture is determined by
\begin{eqnarray}
\partial_{t'}x'&=&E,\nonumber\\
\partial_{y'}x'&=&-B.
\end{eqnarray}
Therefore the coordinate $x_{11}$ plays the role of the coordinate $x'$ in the T-dual picture. Although these two coordinates depend on $E$ and $B$ quite differently, there is one interesting limit ($E\rightarrow1$ and $B\ne0$), where they show very common features: Both M-brane and D-string move at the speed of light and their tilting angles differ by $\pi/2$. This implies that most of the 8-supersymmetric configurations discussed in Sec. \ref{sec5} are U-dual to the intersecting M2-branes moving at the speed of light. Especially the signature of the magnetic field determines the moving direction along the compact $x_{11}$-direction.
The null scissors configuration discussed in Sec. \ref{sec6} is U-dual to two intersecting M2 branes in relative motion. One M2 brane is static but tilted at an angle $\tan{\theta}=E/\sqrt{1-E^2}$ while the other M2 brane is not tilted but moving with the speed $-B/\sqrt{1+B^2}$. We note here that they intersect on a line (but may be displaced by a distance in all other transverse directions) that is moving with the speed $-B/\sqrt{E^2+B^2E^2}=-\mbox{sgn}(B)$, that is, the speed of light. Here, we used the supersymmetric relation (\ref{eb}). Therefore the result of Ref. \cite{acharya} is recovered with intersecting M2 branes. The other 8-supersymmetric configuration discussed in Sec. \ref{sec6} involves a D-string moving at the speed of light. Since its tilting angle is $\pi/2$, BI field strengths are determined to be $E=1$ and $B=0$. Since these values are critical, Eq. (\ref{lll}) becomes ambiguous unless we are given some regulating relation between $E$ and $B$. Hence it is difficult to see its corresponding M2 brane configuration. 

\subsection{Stability and Outlooks}

We would like to mention the issue on the stability of the various 8-supersymmetric configurations discussed in this paper. Although we checked the preserved supersymmetries using the method that resorts to the string theory (rather than low energy effective theories), it is far from clear whether the supersymmetric configurations are stable under small perturbation including the fluctuations about the BI background fields. In order to see this explicitly, we have to check the spectrum of the interstring running between D-branes. As for the intersecting D-strings in uniform motion, this check was already made in the T-dual set up \cite{ohta}. In the paper, it was shown that the tachyonic mode disappears in the supersymmetric limit ($E\rightarrow1$). As for the null scissors configuration, it was argued in \cite{bachas} that a non-BPS excitation of the vacuum state on the one D-string can trigger the joining and splitting interaction between the two D-strings, which could result in the change of the vacuum state on the world volume theory. 

However, as for the other 8-supersymmetric configuration of two intersecting D-strings which are under an infinite relative boost, the stability looks very puzzling. According to the arguments made in \cite{bachas,myers}, the superluminal speed of the intersecting point will cause the interstring either to break off or to pump out energy from two D-strings by stretching against its tension. In both cases, the original configuration will be deformed and the vacuum of the world volume theory will be changed. However, this expectation does not look so natural in the T-dual set-up. Let us consider the configuration composed of two parallel D2-branes shown in the lower picture of Fig. 3. On the one D2-brane carrying the magnetic field only, the world volume theory will be the noncommutative field theory. On the other D2-brane carrying the critical electric field only, the effective world volume theory is not ordinary field theory but the stringy NCOS theory discussed in \cite{seiberg2}. Therefore the full world volume theory will be somewhat `heterotic' $U(1)\times U(1)$ gauge theory which is a noncommutative field theory with respect to the first $U(1)$ factor and is a NCOS theory with respect to the second $U(1)$ factor. Even when the two D2-branes are displaced within the string scale, the pair creation of interstrings (which surely causes the interaction between D2-branes) does not seem to happen. This is because the effective string coupling of the NCOS theory is vanishing; $G_s=g_s\sqrt{1-E^2}\rightarrow0$ as $E\rightarrow1$ \cite{seiberg2}. By channel duality, this pair creation of the interstrings can be understood as the exchange of the closed strings, which is not probable in the vanishing coupling limit. Even when this pair creation happens, the two string-ends, on the D2-brane with the electric field only, will part from each other quickly due to the electric field while the other two ends on the other D2-brane will circulate around within some finite region. The possible decay will be that these circulating two string-ends merge into a long tensionless string ending on the D2-brane with the electric field only. This issue is very important in regard to the recent interests in the time dependent orbifolds and their stability \cite{liu,lawrence,liu2,fabinger,horowitz}. The intersecting D-branes in relative motion will provide a good time dependent background for the string if the configuration is supersymmetric and stable. It is worth while to pursue this issue further.

\begin{acknowledgements}
This work was supported in part by a grant No. R01-2000-000-00021-0 from Korea Science \& Engineering Foundation. We thank Hyeonjoon Shin for helpful discussions.
\end{acknowledgements}

\end{document}